\documentstyle[11pt]{article}

\def\sW{ \sin{\theta_W} }
\def\cW{ \cos{\theta_W} }
\def\Fch{ F^+,~F^-}

\def\beq{ \begin{equation} }
\def\eeq{ \end{equation} }
\def\beqa{\begin{eqnarray} }
\def\eeqa{\end{eqnarray} }
\def\beqa*{ \begin{eqnarray*} }
\def\eeqa*{ \end{eqnarray*} }

\def\PR{ {\it Phys. Rev.} }
\def\PRL{ {\it Phys. Rev. Lett. } }
\def\PL{ {\it Phys. Lett. } }
\def\RMP{ {\it Rev. Mod. Phys.} }
\def\NP{ {\it Nucl. Phys. } }
\def\NC{ {\it Nuovo Cimento} }
\def\etal{{\it et al.}}

\begin{document}
\title{THE FIRST GAUGE THEORY OF WEAK INTERACTIONS AND THE
          PREDICTION OF WEAK NEUTRAL CURRENTS\thanks{Talk given at the
Third International Symposium on the History of Particle Physics:
``The Rise of the Standard Model'', SLAC,  June 24-27, 1992.}}
\author{S.\ A.\ Bludman\thanks{Supported in part by DOE Contract No.
DOE-AC02-76-ERO-3071.}\\
   Department of Physics,   University of Pennsylvania, Philadelphia, PA 19104}

\date{UPR-532-T}

\maketitle

\begin{abstract}
The three theoretical and historical components of the Standard Model are
the exact chiral gauge theory of weak interactions, electroweak unification,
and the Higgs mechanism for spontaneous symmetry breaking.  I put
into historical perspective my
1958 invention of the first gauge theory of weak interactions, predicting weak
neutral currents, and show how the fundamental
differences between global and gauge symmetries
and between partial flavour and exact gauge symmetries, emerged in the strong
and weak interactions.  Although renormalizability is necessary for
theoretical consistency, electroweak unification was not
necessary, in principle.
An interesting difference appears
between the $\sin^2{\theta_W}=0$ limit of the electroweak theory and the
original $SU(2)_W$ gauge theory of weak interactions.
While the electroweak mixing angle
might have had any value (including zero), historically
the value $\sin^2{\theta_W} \sim 0.3$
actually observed in weak neutral currents gave circumstantial support for
the Standard Model and stimulated the search for $W-$ and $Z-$mesons.

\end{abstract}

\section{INTRODUCTION}
The Standard Model \cite{Glashow 1980,Salam 1980,Weinberg 1980}
contains three theoretically and historically distinct elements:
(1) A chiral gauge theory of weak interactions with
an exact $SU(2)_L$ symmetry \cite{Bludman 1958};
(2) The Higgs mechanism \cite{Higgs}
 for spontaneous symmetry, giving some of the gauge bosons finite masses,
while maintaining renormalizabilty \cite{'t Hooft};
(3) Electroweak unification through $W^0-B^0$
mixing by $\sW$ \cite{Glashow 1961}.
Theoretical consistency requires that a field theory be renormalizable, but
not necessarily unified: Within the Standard Model, the electroweak
mixing angle, $\sW$, could in principle have any value, including zero.

In Section 2 of
this report, I recall the history of gauge theories in the 1950's
and my own motivation for publishing \cite{Bludman
1958} the first chiral gauge theory of weak interactions, predicting weak
neutral currents of exact $V-A$ form and approximately the weak strength
observed 15 years later \cite{Hasert}.  In Section 3, I discuss the
historic development of the distinctions
between
global and gauge, partial and exact symmetries, in the weak and strong
interactions. We will see how Goldstone mesons were originally misperceived
as an obstacle to broken-symmetry in the weak interactions.  In Section 4, I
emphasize the theoretical and historic importance of the Higgs mechanism
for symmetry breaking in theories with exact gauge symmetries and the
conditional
role actually played by electroweak unification.  An important property
of the Standard Model in the hypothetical $\sin^2{\theta_W}=0$ limit
emerges,distinguishing it from a purely $SU(2)_W$ theory of weak interactions.
In Section 5, I conclude that, although theoretical consistency did not
require electroweak unification, {\em historically}
the discovery of WNC with electroweak mixing angle
$\sin^2{\theta_W}\sim (0.2-0.3)$
provided circumstantial evidence for the
Standard Model and predicted massive gauge bosons
that were finally observed in 1982.

\section  {NON-ABELIAN GAUGE THEORY FOR THE WEAK INTERACTIONS}

Pauli \cite{Pauli} and the early successes of QED had established the
importance of electromagnetic gauge invariance and how, in simple enough
theories, it led to minimal electromagnetic interactions that were
renormalizable.  For charged vector mesons, however, minimal electromagnetic
interaction was ambiguous \cite{Bludman 1963} and the theory was
non-renormalizable.  The divergences derive from the longitudinal component
of the massive vector meson field and are minimal if the gyromagnetic
ratio $g=2$ and the electric quadrapole moment $Q=-e(\hbar/Mc)^2$.
(In the Standard Model, the electroweak scale acts
as a
regulator for the longitudinal vector meson field, making
vector meson electrodynamics renormalizable for just these electromagnetic
moments.)

I had always been impressed by the Noether's Second Theorem.
While her First Theorem asserted that global symmetries of the Lagrangian
implied well-known conservation laws,
her Second Theorem was much more powerful:
{\em local} Lagrangian symmetries implied {\em new} (gauge) fields.
This, together with the
Yang-Mills theory \cite{Yang}, led to my first publication entitled
``Extended Isotopic Spin Invariance and Meson-Nucleon
Coupling '' \cite{Bludman 1955}, which showed that the then-current pion-
nucleon interaction could not be derived directly from a gauge principle.
Because I was always motivated only by exact gauge symmetries \cite{Kemmer},
I did not think to make the axial current partially conserved, or the
pseudoscalar pion a pseudo-Goldstone meson.  While
approximate flavour $SU(3)$ gauge symmetries led Sakurai \cite{Sakurai}
to hadronic vector mesons, we now realize that only colour
is an exact
hadronic symmetry and that the approximate flavour symmetries derive
from the mass hierarchy of quarks in QCD.

Once the experimental situation clarified in 1957, Sudarshan and
Marshak \cite{Sudarshan}, Feynman and Gell-Mann \cite{Feynman} and Sakurai
\cite{Sakurai 1958} each immediately presented their own derivations of the
$V-A$
$\beta$-decay
interaction.  My own derivation \cite{Bludman 1958} followed from what I
called Fermi gauge invariance, generated by charge-raising and -lowering
chiral Fermi charges $\Fch$.  If the
 algebra of generators is to
close, then neutral Fermi charges $2iF^0=[F^+,~ F^-]$ are
required, i.e. $SU(2)_L$ is the minimal symmetry of the Fermi interactions.
I went on to impose this symmetry locally and was led to an $SU(2)_L$
triplet of gauge bosons, $W^{\pm,~0}$, coupled to a triplet of chiral
Fermi currents ${F_\mu}^{\pm,~0}$.
This chiral gauge theory predicted weak neutral currents of exact $V-A$ form
and the same strength as the weak charged currents.  The observed strength
of the Fermi interactions, $G_F/\sqrt{2}=g^2/8{M_W}$,
then required, in tree approximation,
$M_W=gv/2$, where $v\equiv(\sqrt{2} G_F)^{-1/2}=246
{}~GeV$.  Neither $g$ nor
$M_W$ was predicted separately, but if $g\sim 1$, $M_W \sim 100~GeV$ was to
be expected.

No attempt was made to
provide a mechanism for giving the
intermediate vector bosons mass, to unify weak with electromagnetic
interactions, or to explain the absence of flavour-changing weak neutral
currents.  Flavour-changing WNC were known to be absent to
\( {\cal O}(10^{-8}) \) and even flavour-preserving WNC were
incorrectly reported
\cite{Block} to be at least thirty times weaker than charged weak neutral
currents.  Faith in quarks and in quark-lepton symmetry
soon led Glashow \etal
\cite{GIM} to propose the GIM mechanism, explaining the absence of
flavour-changing WNC at tree-level and reducing their radiatively-induced
amplitude at
${\cal O}(G_F\alpha)$
 by a suitably large factor
$({m_c}^{2} -{m_u}^{2})/{M_W}^{2}.$

My 1958 paper was soon followed by proposals \cite{Lee and Yang 1960,%
Pontecorvo,Gershtein} to use accelerator neutrino beams to search for
flavour-preserving weak neutral currents.
But this search remained very difficult, because of high backgrounds for
 neutrino-
induced charged-current processes in which muons escaped undetected which
were hard to estimate. In any case, neutrino experimentalists in the 1960's
were preoccupied with deep inelastic scattering at SLAC and scaling.

These experimental difficulties, together with the
need for a consistent theory allowing massive gauge bosons,
suggest why chiral weak neutral currents needed to wait from 1958 to 1973
for experimental confirmation.

\section {SPONTANEOUSLY BROKEN GLOBAL SYMMETRIES}

The idea of spontaneous symmetry breaking (SSB), better denominated
hidden symmetry, was brought from condensed matter physics to quantum field
theory by Heisenberg \cite{Heisenberg} and Nambu \cite{Nambu} and soon led
to the Goldstone Theorem \cite{Goldstone,Klein}. Klein and I identified
the Goldstone bosons expected from different levels of global symmetry-
breaking and emphasized that Goldstone bosons were not present in
theories with long-range interactions.  The Goldstone Theorem showed how
SSB could produce long-range interactions out of a short-range theory.
Following Anderson \cite{Anderson},
we suggested that, conversely, long-range
interactions might be converted into short-range.  But we missed the Higgs
mechanism which differentiates between
 the role of Goldstone mesons in gauge theories
and their role in global symmetry theories.

In the weak interactions, Klein and I observed there
were apparently no Goldstone bosons and that,
although the neutrino was apparently massless, it could not be a Goldstone
meson, because the vacuum could not be macroscopically occupied by fermions.
Because
we failed to associate the Goldstone Theorem with my earlier
proposal of a
{\em gauge} theory of weak interactions,  Goldstone mesons were misperceived as
obstacles to a
theory of
weak interactions,

The 1958 work on chiral invariance was cited
 by Gell-Mann and by Nambu
\cite{Gell-Mann 1960} and ultimately led to current algebras, soft-pion
theorems, PCAC and the Goldberger-Treiman relation.
These heuristic successes, however, tended
to gloss over the fundamental differences between global and gauge
symmetries, and between
partial
flavour symmetries and exact gauge symmetries, in the strong and weak
interactions.

Exact symmetries were useful in
classifying fields and particles and were most aesthetically satisfying.
For these reasons, I tended to avoid
hadron physics and concentrated on weak interactions where, I was
convinced, exact symmetries were to be found.
At his time, I left the University of California Radiation
Laboratory (now the Lawrence Berkeley Laboratory), which was then dominated
by dispersion relations and S-matrix theory.
I took an academic position at the University of
Pennsylvania and my interests gradually shifted from laboratory to
astrophysical particle physics.

\section {SPONTANEOUSLY BROKEN GAUGE SYMMETRIES, WITH AND WITHOUT UNIFICATION}

Unlike earlier global theories of the strong interactions \cite{Kemmer},
my 1958 theory of the weak interactions was an exact (chiral) gauge theory.
The power< of an exact gauge symmetry is that, if symmetry is spontaneusly
broken by the Higgs mechanism \cite{Anderson,Higgs}, giving some gauge bosons
masses,
the symmetry remains hidden and the theory remains renormalizable.  Indeed,
the Standard Model has only exact gauge symmetries,
massless gauge bosons
are usually not manifest: Either the
(colour) gauge symmetry is unbroken, but the massless gluons
are confined, or the gauge symmetry is spontaneously
broken,  providing masses for
the gauge bosons.

Weinberg \cite{Weinberg 1967}
and Salam \cite{Salam 1968} proposed the Electroweak Standard Model,
conjecturing that the theory would remain renormalizable.
Nevertheless,
the 1967 Weinberg paper was referred to by no one (including Weinberg) in
1967-70 and only once in 1971 \cite{Coleman}.
Finally,  `t Hooft \cite{'t Hooft} proved
that such a theory remained  renormalizable.   In this way, a complete
theory
of massive charged vector mesons and of
weak interactions was achieved \cite{Glashow 1980,Salam 1980,Weinberg 1980}.

In the minimal Standard Model, the couplings enter through the
$SU(2)_L\times U(1)_Y$ covariant derivative
$D_\mu=\partial_\mu - ig{\bf T}\cdot{\bf W}_\mu -ig'(Y/2)B_\mu.$

(1) The Higgs mechanism gives the vector mesons (generally) unequal
masses $M_W=M_Z~\cW,$ where $g'/g \equiv \tan {\theta_W}.$
(2) The charged vector mesons couple to the electromagnetic field with magnetic
moment $2(e\hbar/2Mc)$ and electric quadrapole moment $-(\hbar/Mc)^2$;
(3) the WNC couple to $Z^0$ with coupling constant $g/\cW$;
(4) Charged particle currents couple to the electromagnetic field with
coupling constant
$e \equiv g~\sW \equiv g'~\cW.$  Consequently, in tree approximation,
\beq
M_W \sW=M_Z\sW\cW=(e/2)(\sqrt{2}G_F)^{-1/2}=\sqrt{\pi\alpha} v\equiv A_0=
37.3~GeV.
\eeq

For consistency, a quantum field theory needs to be renormalizable, but
need not be unified.
In principle, we could have had either
spontaneously broken  $U(1)$ symmetry with no weak currents
($g=0,~g'=e,$ Schwinger's
electrodynamics with massive photons \cite{Schwinger 1962}) or weak $SU(2)_L$
symmetry ($g'=0=e$, WNC with coupling constant $g$ \cite{Bludman 1958}).
These two examples illustrate, contrary to ref. [1],  the logical possibility
of consistent (renormalizable) theories without unification.

The electromagnetic field exists and $B^0$ and $W^0$ mix, but
how Nature chooses  $\sin^2{\theta_W}=0.23~~(e \leq g' \leq g < 1),$
remains unexplained within the
Standard Model. In the minimal Electroweak Model,
besides $G_F$ and
$e$, there is only one free parameter,
\beq
\sin^2{\theta_W} \equiv
(M_Z^2 - M_W^2)/M_Z^2=(1/2)[1-\sqrt{1-(2A_0/M_Z)^2],}
\eeq
in tree approximation, which measures the $SU(2)_L$
symmetry breaking through $W^0 -B^0$ mixing.
In a unified theory, $g,~g' \geq e$, so that $M_W \geq 37~GeV, ~M_Z \geq
74~GeV.$  This leads to an interesting difference between a pure $SU(2)_W$
theory and the $\sin^2{\theta_W} \rightarrow 0$ limit of the
Electroweak Theory.  In
the former, only the ratio $M_W/g=123~GeV$ is constrained.  In the
latter, holding $G_F, ~e$ constant as $\sin^2{\theta_W} \rightarrow 0$,
makes $g' \rightarrow e$ and
$g,~M_W=M_Z$ diverge.  The $\sin^2{\theta_W} = 0$
limit of a
unified theory would be one with tree-approximation point-interactions, which
is unitary and renormalizable because of huge radiative corrections!

For energies $\gg 10~GeV,$ the mass differences between W- and Z-mesons and
among the quarks (other than the top quark) can be neglected and
the original $SU(2)_L$ symmetry is restored.  This old theory will
then be
 a good approximation to leptonic and semi-leptonic processes, other than
top quark decay.  Thus, the effects
of unification practically disappear already
at energies $\gg 10~GeV$ much lower
than the unification scale at which symmetry-breaking dissappears.

\section {HISTORICAL CONCLUSIONS}

The proof that, in an exact gauge theory, renormalizability would be
retained even as the gauge mesons acquired mass by the Higgs mechanism,
immediately convinced
 theorists and experimentalists of the Electroweak Standard Model.
Although a consistent theory without
electroweak is possible, in principle,
the discovery of weak neutral currents \cite{Hasert} with
mixing $\sin^2{\theta_W}\sim 0.3,$
gave circumstantial evidence for the Standard Model and predicted $M_W
\approx 80~GeV, ~M_Z \approx 90~GeV$.  The ultimate discovery
\cite{Discover} of these gauge bosons with unequal masses
directly confirmed the Standard Model.

\end{document}